Physics of the Space Environment

# Spacecraft Charging and Hazards to Electronics in Space

Tsoline Mikaelian
York University

May 2001

## --- ABSTRACT ---

The interaction of a space system with its orbital environment is a major consideration in the design of any space system, since a variety of hazards are associated with the operation of spacecraft in the harsh space environment. In this brief review, two types of hazards to Earth-orbiting spacecraft are discussed: spacecraft charging and radiation hazards to spacecraft electronics, with emphasis on the natural environmental factors and interactions which contribute to these hazards. Following a summary of the historical eras of spacecraft charging and some observations from experimental satellites: SCATHA, CRRES and DMSP, environmental factors significant to spacecraft charging are discussed, including plasma interactions, electric and magnetic fields and solar radiation. Spacecraft charging depends on the spacecraft geometry, as well as on the characteristics of its orbit, since the natural environment may differ for each type of orbit. Low altitude orbiting satellites (LEO) usually experience less charging effects than high altitude geosynchronous (GEO) satellites, except for low altitude polar orbiting satellites which cross the auroral oval. Basic mechanisms of surface charging, differential charging and internal charging are described. Environmental factors including trapped and transient radiation, solar and galactic cosmic rays, which can profoundly damage spacecraft electronics are presented. Some effects such as ionization and atomic displacement damages to semiconductors and single event phenomena are also briefly mentioned.

## CONTENTS

# I. INTRODUCTION

The space environment has a complex and dynamic structure. It includes neutral species, charged particles, plasmas, electric and magnetic fields, solar and galactic radiation, meteoroids and space debris, each of which can cause profound damages to a spacecraft, deteriorating its performance and lifetime. Interactions between a space system and its environment can cause modifications to the natural environment, giving rise to local environments which in turn affect the system's behavior (Purvis, 1993). Numerous operational anomalies and satellite failures have been reported since the beginnings of the "space age", a significant number of which were attributed to the phenomenon of spacecraft charging and to radiation effects on electronic systems. In the first half of this paper, emphasis is placed on spacecraft interactions with environmental factors which give rise to spacecraft charging, along with associated effects. In the second half of the paper, the space radiation environment is presented and some of its damaging effects on electronic systems are described.

In the context of this paper, the space environment refers to the near Earth space environment, which spans a range from low Earth orbits (LEO) below around 1000 km altitude, to beyond geosynchronous orbits (GEO) at around 35,000 km altitude.

Spacecraft charging is an important consideration in the design of spacecraft, since it can have critical effects on spacecraft operation. The NASA reference publication 1375 (Leach and Alexander, 1995) lists some cases of the operational anomalies caused by spacecraft charging. Considerable efforts have been made to enhance our knowledge of the basic mechanisms involved, and to develop theories and simulation techniques to predict and prevent charging effects. Also, radiation effects on electronics have been extensively studied since the early predictions and observations of anomalies in space electronic devices. Nowadays, with the utilization of sophisticated electronic circuits, which operate at low voltages and low currents, the effects of spacecraft charging and radiation on electronics have become increasingly important. Section II of this review is devoted to a brief historical overview of spacecraft charging and radiation effects and the remarkable missions of the SCATHA, CRRES and DMSP satellites, which represent milestones in the study of space radiation effects and charging phenomena.

Following the historical review, an overview of spacecraft charging environments is made in section III. Spacecraft charging can occur in the form of surface charging or internal charging. Surface charging is defined as the accumulation of charge on the spacecraft surface. Plasma interactions, charged particles, solar radiation and magnetic fields are the major contributors to surface charging. Internal charging, on the other hand, occurs due to highly energetic electrons, which can penetrate deep into dielectric materials inside the spacecraft. Both surface and internal charging environments are described, along with the basic charging mechanisms and associated effects. The GEO and LEO environments are also described and the special case of low altitude polar orbits is noted. Various anomalies attributable to spacecraft charging are presented. Finally, a brief reference to computer simulations and design guidelines is made. In this review, an effort has been made to give a general overview of the broad subject of spacecraft charging which has been subject to continuous studies for over 30 years. This review lacks detailed mathematical treatments of charging theory, models and simulations. For in-depth review of spacecraft charging, the reader is referred to a paper by Garrett (1981).

Spacecraft electronics are subject to space radiation hazards. The space radiation environment includes trapped electrons and protons of the Van Allen radiation belts, and non-trapped, transient solar and galactic cosmic rays and solar flare particles. The South Atlantic Anomaly (SAA) contributes greatly to the radiation environment and is also briefly discussed. Several regions of trapped radiation are identified. Finally, some basic radiation effects on electronics are presented, including ionization and atomic displacement effects in semiconductors, and single event phenomena. However, mechanisms of the space radiation effects on various electronic devices are not discussed in this review. The reader is referred to the excellent review papers of Srour and McGarrity (1988) and Stephen (1993) for discussions of radiation effect mechanisms.

## II. HISTORY: SPACECRAFT CHARGING AND RADIATION EFFECTS

Garrett subdivided the developments in spacecraft charging into five different historic periods (Garrett, 1981; Garrett and Whittlesey, 1996). The first period of spacecraft charging research began with the electrostatic probe work of Langmuir (Langmuir and Blodgett, 1924; Mott-Smith and Langmuir, 1926). During this period, much of the interest was in the potentials of space dust particles. Spacecraft charging emerged as a discipline in the 1950's, as soon as rockets equipped with sensors were developed and used for ionospheric measurements. The first effects of spacecraft charging were reported in 1955 (Johnson and Meadows, 1955).

The launch of Sputnik in 1957 denoted the beginning of the second period in spacecraft charging advances. In 1961, the first review paper on spacecraft charging appeared (Chopra, 1961) and a significant portion of spacecraft charging theory was established. Sputnik 3 was the first satellite to make potential measurements. Rocket and satellite observations confirmed that charging existed and presented dangers to spacecraft operations.

The third period was marked by developments of realistic models of spacecraft charging, along with accurate measurements by rockets and satellites. The first complete book concerned with spacecraft charging appeared in 1965 (Singer, 1965). Reviews by Brundin (1963) and Bourdeau (1963) and Whipple's thesis (Whipple, 1965) are comprehensive works of this period.

The fourth period, extending from 1965 to 1980, was characterized by very sophisticated theories of spacecraft charging and spacecraft interactions with plasma environments. In 1973, a catastrophic failure of the US Air Force Defense Space System

Communication Satellite (DSCS) 9431 occurred due to a discharge which resulted in the loss of power to the satellite's communication system. This event lead to the cooperation of NASA and the Air Force to develop technologies to control spacecraft charging. Computer codes such as NASCAP (NASA Charging Analyzer Program) were developed to predict charging effects (Katz et al., 1979; Roche and Purvis, 1979). Design guidelines were also documented in NASA Technical Report 2361 (Purvis et al., 1984) and MIL-STD 1541A (Anon., 1987).

The end of the fourth period was denoted by the 1979 launch of SCATHA (Spacecraft Charging at High Altitude) satellite, also known as P78-2. SCATHA's main mission was to obtain information about spacecraft surface charging effects and processes. SCATHA successfully gathered environmental and engineering data, which allowed the development of methods to control charging. It also collected scientific data for the study of plasma wave interactions and substorms (Mullen et al., 1981).

The period from 1980 to the present marks the fifth phase of spacecraft charging. This period is characterized by an increased awareness of internal charging phenomena and Low Earth Orbit (LEO) charging effects. Since the early launches of LEO satellites and the initial plans of deploying the International Space Station, research in low altitude charging environments was emphasized. In 1982 and 1983, two DMSP (Defense Meteorological Satellite Program) satellites were launched into polar orbit to study the low altitude polar environments and gather information about auroral activities (Daly and Rodgers, 1993; Gussenhoven, 1985).

Spacecraft charging is hazardous to onboard electronics. In addition, energetic particles of the space environment can impact semiconductor devices and lead to transient upsets. The first prediction of such an upset (called single event upset) was made in 1962 (Wallmark and Marcus, 1962). The first upsets were observed onboard Intelsat IV in 1975 (Binder et al., 1975). Following these observations, several groups indulged in the study of cosmic ray effects on semiconductors, as well as a thorough study of the natural space radiation environment.

Nowadays, very small integrated circuits are being extensively used aboard satellites and these microelectronic circuits are very sensitive to charging and radiation effects. A sound understanding of the radiation environment and its effects on microelectronics was necessitated with the rapid technological advances in microelectronic circuits. The CRRES (Combined Release and Radiation Effects Satellite) spacecraft, which was launched in 1990, studied the natural radiation environment and its effects on microelectronics. CRRES performed a series of internal charging experiments (Fredrickson et al., 1992). It traveled through the Earth's radiation belts, exposing microelectronics to the radiation environment and at the same time accurately mapping the radiation belts. Such experiments allowed engineers to establish a correlation between microelectronic performance and radiation exposure levels.

This review is concerned with the space environment and its basic interaction with spacecraft as relevant to spacecraft charging and radiation effects. Surface charging in geosynchronous orbit is discussed, as well as low altitude charging, internal charging and hazards to electronics, which have become significantly important and have been subject to extensive studies for decades.

## III. SPACECRAFT CHARGING

Spacecraft charging is defined as the buildup of charge on spacecraft surfaces or in the spacecraft interior. Spacecraft charging causes variations in the electrostatic potential of a spacecraft surface with respect to the surrounding plasma environment, and/or potential variations among different portions of the spacecraft (differential charging). The major natural space environments which contribute to spacecraft charging include the thermal plasma environment, high energy electrons, solar radiation and magnetic fields, each of which is discussed in this section. Spacecraft charging has many effects, the most dangerous of which is perhaps electrostatic discharges which can have catastrophic consequences such as structural damage, degradation of spacecraft components and operational anomalies caused by damages to electronics.

In this section, an overview is presented of the space plasma environments and other environmental factors relevant to spacecraft charging, different types of charging phenomena, and effects of spacecraft charging.

**1. Spacecraft Charging Environments**

*a. Plasmas*

The Earth's atmosphere is homogeneous up to an altitude of approximately 90 km. Above this altitude, a significant portion of the atmospheric molecules are ionized due to solar radiation, producing free electrons and positively charged ions. An ionized gas containing equal numbers of positively and negatively charged particles is known as plasma. The ionized gases of the atmosphere constitute the natural space plasma environment, which interacts with all spacecraft in near-Earth orbit (including low altitude as well as geosynchronous orbits of all inclinations).

The plasma environment varies with altitude and latitude. Plasma environments relevant to spacecraft charging that are discussed in this paper are the LEO, GEO and polar environments. Low altitude, low inclination spacecraft will experience a different plasma environment than high altitude geosynchronous spacecraft and polar orbiting spacecraft. Figure 1, taken from NASA Reference Publication 1375 (Leach and Alexander, 1995), illustrates the particle density and energy characteristics of the natural space plasma in LEO, GEO and polar orbits. As obvious from Figure 1, the two major types of plasmas are the low energy/high density plasmas and the high energy/low density plasmas.

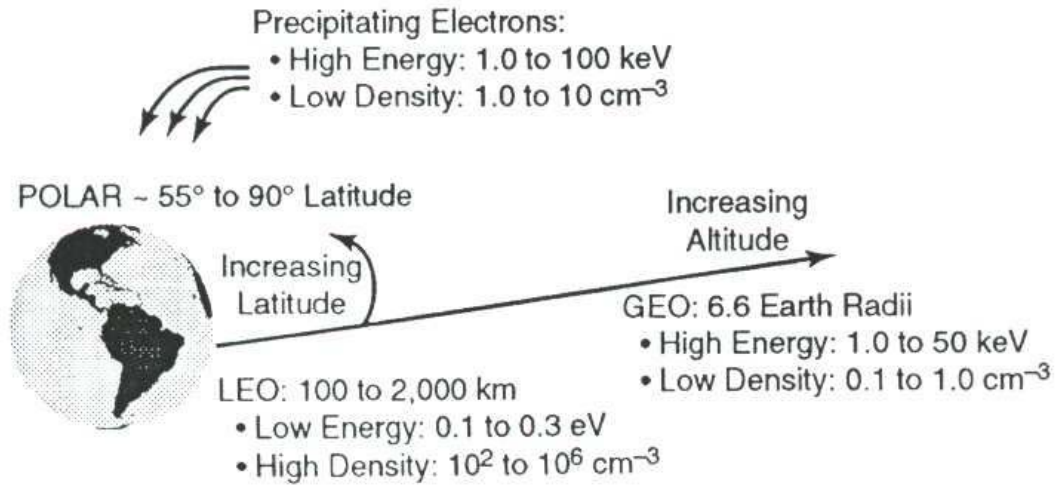

**Figure 1.** Properties of the natural space plasma *(after NASA RP 1375)*.

The low energy/high density plasma found in LEO is the ionospheric plasma. The ionospheric plasma is overall neutral and has characteristic temperature in the order of 0.1 eV. The peak density of ionospheric plasma is around $10^6$ /cm$^3$ at about 300 km magnetic equatorial latitude, as shown in Figure 2 (Purvis, 1993). The ionospheric plasma co-rotates with the Earth's magnetic field and its density also varies with latitude, being greater at magnetic equatorial latitudes than at polar latitudes.

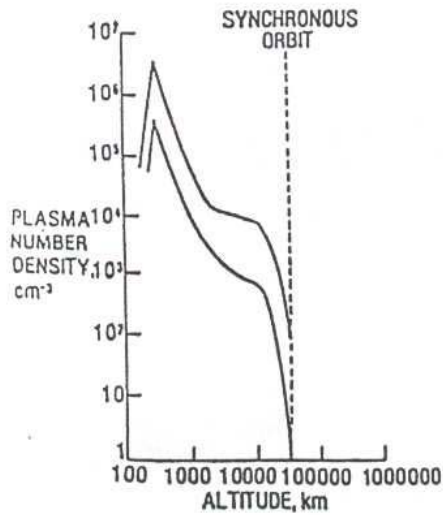

**Figure 2.** Variation of plasma density with altitude *(after Purvis, 1993)*.

The high energy/low density plasma found at GEO and polar environments is associated with geomagnetic substorm activity. These plasmas have temperatures in the order of tens of keVs, and a density less than 1.0 /cm$^3$. They are carried down magnetic field lines and produce high-energy particle streams which precipitate in the polar auroral zones. During geomagnetic substorms, the flux of high-energy electrons accelerated towards the Earth increases.

For the purpose of analyzing spacecraft interactions with the surrounding plasma environment, the LEO ionospheric plasmas and the GEO plasmas associated with magnetic substorms must be treated differently. This is because depending on the characteristics of each plasma, potentials on spacecraft are screened from their surroundings to varying degrees. A *sheath* is the region of space surrounding a body, which feels a perturbation of potential caused by that body. If a small source of potential perturbation $V_o$ exists in the plasma, the potential surrounding it falls off as

$$V = V_o e^{-\frac{|r|}{\lambda_D}} \qquad (1)$$

where $\lambda_D$ is called the Debye length. The Debye length is a scale length for the size of the sheath, i.e. a characteristic length over which the potential falls off by a factor of 1/e (Chen, 1974). It is given by

$$\lambda_D = \sqrt{\frac{\varepsilon_o K T_e}{e^2 n}} \qquad (2)$$

where $\varepsilon_o$ is the permittivity of free space, K is Boltzmann's constant, $T_e$ is the electron temperature (in Kelvin), e is the electron charge and n is the ambient plasma density. Equation (2) indicates that the Debye length varies with the electron temperature and plasma density, which are the quantities that vary for each of the regions discussed above. Therefore, in low altitude and latitude orbits, the relatively cool and dense ionospheric plasma has short Debye length, in the order of millimeters. This plasma region is thus referred to as a *thin sheath regime*. In contrast, in the high altitude regions, the substorm plasma is characterized by large Debye length, and forms a *thick sheath regime*. Note that equations (1) and (2) apply only to situations where small perturbing potentials are involved and hence Poisson's equation can be linearized. The case of high level charging may be analyzed using the non-linear theory developed by Child and Langmuir (Katz, 1993). The concepts of Debye length and sheath thickness are revisited in subsequent subsections, where discussions of spacecraft charging in various orbits are presented.

The plasma affects spacecraft by inducing charges on the spacecraft surface due to the flux of electrons and positive ions. The motion of a spacecraft through plasma may give rise to a local environment which may also contribute to spacecraft charging. This section summarized some basic background on plasmas, necessary for the study of the charging phenomenon.

*b. High Energy Electrons*

Auroral regions and the GEO environments include high-energy electron populations which typically have a Maxwellian distribution and a characteristic temperature $T_e$. When a significant plasma environment and photoelectrons arising from solar radiation are not present, the potential to which a spacecraft will charge is directly proportional to the electron temperature and varies between 1 to 20 kV. Electrons with energies between 1-100 keV contribute to surface charging, while trapped electrons with energies above 100 keV penetrate the surface and contribute to internal charging effects.

### c. Solar Radiation

Photons emitted from the Sun have an important effect in surface charging. UV and EUV photon impacts on spacecraft surfaces result in the emission of photoelectrons (by the photoelectric effect). These photoelectrons constitute a current out of the spacecraft surface, which can reduce the effect of negative surface charging and hence it can be an important contributor to the surface charging mechanism described in section 2. The effect of solar photons is particularly important for GEO orbits where the plasma density is low and the contribution of photoelectrons is not negligible. The photoelectron current depends on the surface material of the spacecraft, the solar activity, solar incidence angle and spacecraft potential (Lucas, 1973).

### d. Magnetic Fields

The Earth's magnetic field is approximately a magnetic dipole which is displaced from the center of the Earth by ~ 436 km. The geomagnetic axis is inclined at 11.5° with respect to the rotational axis of the Earth. The Earth's magnetic field has great influence on plasma motions and on trapped high-energy charged particles, which lead to spacecraft charging and damages to electronics. The magnetic field determines the regions of the space environment where spacecraft charging can occur. It also plays a role in the surface charging mechanism since it can affect the escape of electrons (such as photoelectrons) emitted from the spacecraft surface. This idea is illustrated in Figure 3 (Laframboise, 1983). If the magnetic field is nearly perpendicular to the spacecraft surface, electrons can escape along the field lines (Figure 3(a)). On the other hand, if the magnetic field is inclined with respect to the surface, electrons may be redirected to the surface by the Lorentz force law: $\vec{F} = -e\vec{E} \times \vec{B}$ (Figure 3(b)). This prevention of electron escape from the surface increases the negative surface charge.

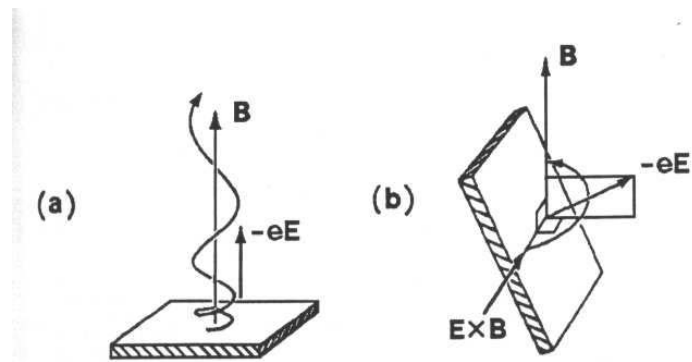

**Figure 3.** Magnetic field effect on secondary electron emission *(after Laframboise, 1983)*.

The passage of a spacecraft through the Earth's magnetic field also induces potentials across the spacecraft, since the motion through a magnetic field gives rise to an electric field. The direction of the induced potential is perpendicular to both the magnetic field direction and the spacecraft velocity vector, and its magnitude V is given by

$$V = (\vec{v} \times \vec{B}) \cdot \vec{L} \qquad (3)$$

where v is the spacecraft velocity, B is the magnetic field, and L is the spacecraft dimension. A typical value of V/L is ~ 0.3 V/m for a spacecraft in LEO, where both v and B are large (Purvis, 1993). At the altitudes of GEO, the induced potential due to spacecraft motion is zero.

## 2. Surface Charging

### a. Mechanism

Surface charging is caused by the interaction of spacecraft surfaces with the plasma environment, solar radiation, high-energy electrons and magnetic fields. A space system moving through the space plasma environment reaches electrical equilibrium with the plasma by acquiring surface charges such that the net current to the whole system and to the individual insulating surfaces is zero (Whipple, 1918; Katz, 1993). This equilibrium condition determines the spacecraft's surface potential relative to the surrounding plasma. A spacecraft surface may consist of conducting and/or insulating materials. For the conducting surfaces, equilibrium is established globally, while for insulating materials the equilibrium occurs on a point-to-point basis. In establishing equilibrium, all natural environmental factors must be taken into account, as well as motionally induced potentials and other voltages generated by the spacecraft itself. The suppression of electron escape from the surface by magnetic fields must also be taken into consideration whenever appropriate.

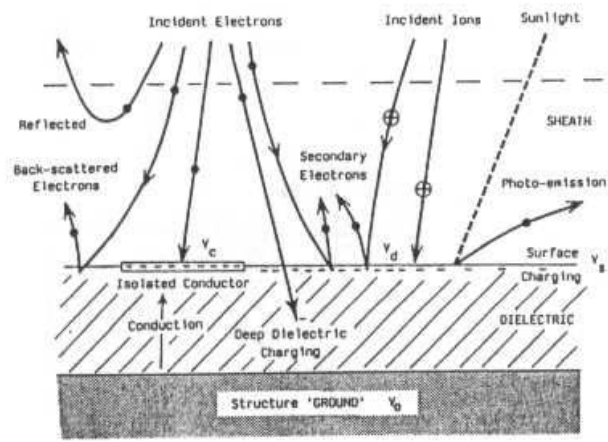

**Figure 4.** Currents which control surface charging
*(after Wrenn and Sims, 1993).*

As an example, Figure 4 illustrates the current densities into and out of a simple isolated insulating surface or conductor. As electrons and positive ions hit the insulating material, secondary electron and ion emissions are produced from the surface. Some back-scattered electrons are also produced from the impact. If the surface is sunlit, photoelectrons are emitted too. Also, leakage currents through the insulator may result. Ignoring leakage currents, the current balance equation is given by:

$$I_{NET}(V) = I_E(V) - [(I_I(V) + I_{SE}(V) + I_{SI}(V) + I_{BSE}(V) + I_{PH}(V)] \tag{4}$$

where:

V = surface potential relative to the plasma; all currents are functions of surface potential.
$I_{NET}$ = total current from the spacecraft surface
$I_E$ = incident environmental electron current
$I_I$ = incident environmental positive ion current
$I_{SE}$ = secondary emitted electron current due to $I_E$
$I_{SI}$ = secondary emitted electron current due to $I_I$
$I_{BSE}$ = back-scattered electron current sue to $I_E$
$I_{PH}$ = photoelectron current due to sunlight

Note that $I_I$, $I_{SE}$, $I_{SI}$, $I_{BSE}$, and $I_{PH}$ act as positive currents (to the surface), since they result either from electrons leaving the surface or positive ions incident on the surface. On the other hand, $I_E$ acts as a negative current since it results from electrons incident on the surface.

The equilibrium condition is therefore given by

$$I_{NET}(V) = 0 \ . \tag{5}$$

At equilibrium, the charging process stops and the spacecraft reaches the equilibrium charging level, also called the "floating potential". Of course, the equilibrium is dynamic in the sense that it changes whenever the current densities change.

The electron and positive ion temperatures and densities are nearly equal in the space plasma environment. However, the ions are considerably more massive than the electrons. This implies that electrons are more mobile than ions and hence the electron flux dominates and the negative current is larger than the positive current. Consequently, the surface charges negatively to a potential in the order of the electron temperature.

### b. *Types: Absolute and Differential Charging*

Absolute surface charging is the charging of the whole spacecraft surface to a net potential relative to the surrounding plasma. If the surface is made up only of conducting materials, the spacecraft charges to the same potential everywhere since the charge distributes uniformly on a conducting surface.

The other type of surface charging is differential charging, which occurs when the spacecraft surface has dielectric materials (such as Kapton$^{TM}$ or Teflon$^{TM}$). Different portions of the spacecraft surface charge to different "floating" potentials in this case. As mentioned earlier, the effect of solar radiation is considerable in GEO where the plasma density is low. Photoelectrons will be emitted from sunlit surfaces and will tend to cancel the effect of the electron current, maintaining the sunlit surfaces at zero potential. On the other hand, in the shaded portions of the surface, photoelectrons are absent and the surface charges negatively. As the negative charge on the shaded surfaces increases, it prevents the emission of photoelectrons from the sunlit surfaces and the entire spacecraft begins to charge negatively. This gives rise to differential potentials in the order of kilovolts across different portions of the spacecraft surface (Purvis, 1983).

Differential charging is more dangerous than absolute charging because it can lead to surface arcing or electrostatic discharges (ESD) among surfaces having different potentials, which in turn give rise to various operational anomalies (Olsen et al., 1981).

### c. *Surface Charging in GEO*

For many years, surface charging of spacecraft had been associated with geosynchronous and nearby high altitude orbits. Wrenn and Sims summarized three basic observations related to GEO spacecraft charging in their review paper (Wrenn and Sims, 1993):

(i) Geosynchronous spacecraft can become charged to more than 10,000 V in magnitude (DeForest, 1972).
(ii) Geomagnetic activity is closely related to the charging anomalies in GEO.
(iii) Spacecraft charging anomalies in GEO occur mostly in the midnight to dawn sector, i.e. they are not uniformly distributed in local time. Figure 5 illustrates this point. This observation can be explained by the fact that during magnetic substorms, electrons are injected from the local midnight region and drift towards dawn.

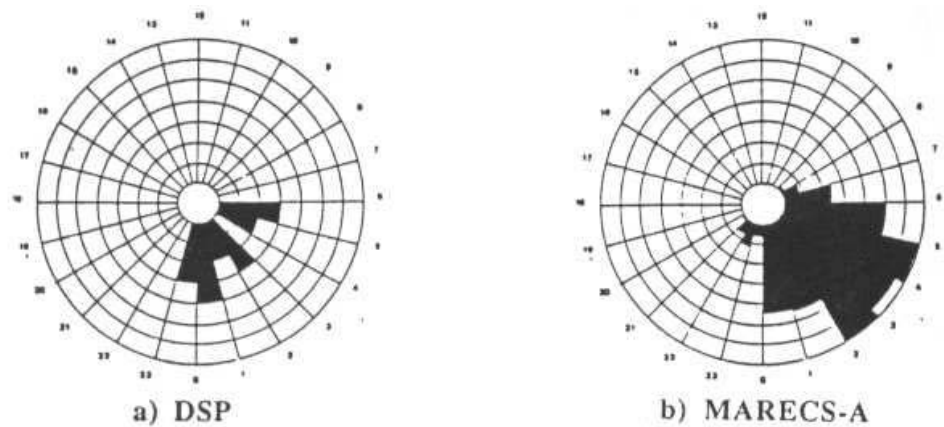

a) DSP    b) MARECS-A

**Figure 5.** Chronograms of events/local time for the DSP and MARECS-A satellites, illustrating event occurrence in the midnight to dawn sector *(after Romero and Levy, 1993)*.

Recall from an earlier discussion that the GEO environment consists of plasma associated with geomagnetic substorms. This plasma has a relatively low density and high temperature and hence the Debye length is very large: in the order of hundreds of meters (thick sheath region). This plasma environment has several implications regarding the analysis of spacecraft charging in GEO:

- Since the plasma is at a high temperature, the spacecraft can be considered to be at rest relative to the surrounding plasma, thus there is no need to study motional spacecraft/plasma interactions.
- Since the plasma is at a low density, photoelectron emission by solar UV/EUV radiation plays an important role in the current balance equation. Considerable differential surface charging may result.
- Since Debye lengths are large, the spacecraft size can be considered to be relatively small compared to the size of the sheath.
- Space charge effects can be ignored in this region, allowing the use of Laplace's equation $\vec{\nabla}^2 V = 0$ for calculating the potential around the spacecraft.
- The potetntials induced on the spacecraft by the natural space environment exceed the voltages generated by the system itself.

### d. Surface Charging in LEO

Two cases can be distinguished for spacecraft charging in LEO. First, there is the case of low latitude orbits which experience the ionospheric plasma. This environment is benign relative to the GEO environment with respect to charging. Second, there is the case of high latitude polar orbits which are subject to precipitating electron streams that vary with auroral activity. While significant charging cannot occur in low latitude ionospheric plasmas, the polar LEO environment can lead to significant charging.

Again, recall from earlier discussions, that the ionospheric plasma is relatively cold and dense, hence small Debye lengths (~ mm or cm) mean that LEO is a thin sheath region, i.e. spacecraft potentials are screened more effectively from the plasma surroundings. Some implications of these plasma characteristics are listed here:

- Since the plasma is relatively cold, the effects of spacecraft motion through the plasma become very important. A discussion of motional spacecraft/plasma interactions is included later in this section.
- Since the plasma is at a high density, photoelectron emission by solar UV/EUV radiation is negligible.
- Since Debye lengths are small, the spacecraft size is relatively large compared to the size of the sheath.
- Space charge effects cannot be ignored in this region, necessitating the use of Poisson's equation $\vec{\nabla}^2 V = -\frac{\rho}{\varepsilon_o}$ (where $\rho$ is the charge density) for calculating the potential around the spacecraft.
- In the cold ionospheric plasma, there is no possibility of high level surface charging. Potentials induced by the spacecraft are comparable to the plasma temperature and thus cannot be ignored.

It is convenient first to give some definitions related to the motion of a spacecraft through plasma. A spacecraft's velocity is *mesothermal* at low latitudes. Mesothermal is defined by the relation

$$V_I < V_{S/C} < V_E \qquad (6)$$

where $V_I$ is the thermal speed of the ions, $V_{S/C}$ is the spacecraft orbital speed, and $V_E$ is the thermal speed of the electrons (Daly and Rodgers, 1993). Typical values are 7.5-8 km/s for $V_{S/C}$, 1 km/s for $V_I$ and 150 km/s for $V_E$. Another important quantity to be considered is the ion sound speed $V_S$. At low altitudes, the ion sound speed is the same as the ion thermal speed $V_I$ (Daly and Rodgers, 1993). Therefore, the spacecraft motion is also *supersonic* with respect to the ions. The spacecraft *Mach number* M is defined as

$$M = \frac{V_{S/C}}{V_S} \qquad (7)$$

Typical values of spacecraft Mach numbers range between 6 and 8.

As a result of its supersonic motion, a spacecraft will create a large volume of disturbed plasma as it moves through its orbit. A region of space depleted from ions will form behind the spacecraft (i.e. opposite to the spacecraft velocity vector). This region is termed *near wake* region. Density enhancement will occur in the ram region in front of the spacecraft, along the direction of motion. Due to these perturbations, electrons will be free to strike all surfaces, while ions can impact only ram surfaces. The hypersonic motion of the spacecraft also gives rise to a shock wave in front and to the side of the spacecraft. The spacecraft/plasma motional interactions are illustrated in Figure 6. For detailed discussions of these interactions the reader is referred to the papers by Daly and Rodgers (1993), Stone (1981), and Wright (1988).

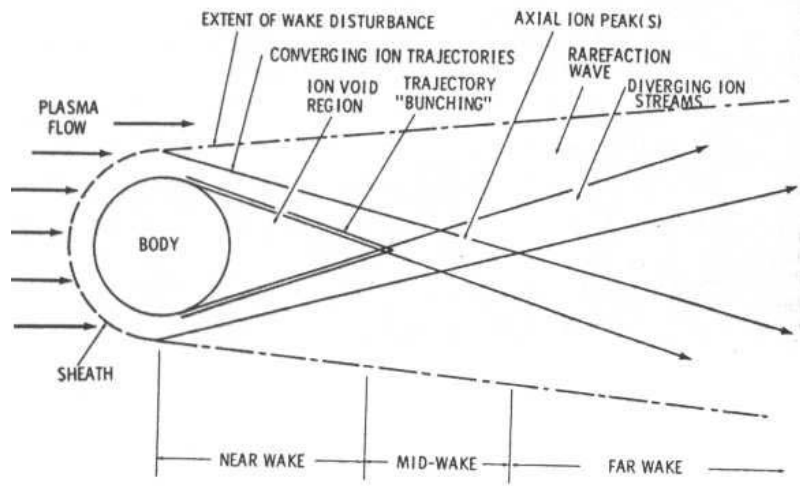

**Figure 6.** Spacecraft/plasma motional interaction in LEO *(after Stone, 1981).*

The special case of LEO spacecraft in high inclination (polar) orbits is an intermediate case between LEO and GEO charging. In polar orbits the interaction of spacecraft with the aurorae can lead to high levels of surface charging. Characteristic lengths and potentials vary depending on the orbit position and auroral activity. Inhibition of electron escape from surfaces by magnetic fields also plays an important role in the polar regions. Figure 7 shows plasma interactions with a spacecraft in polar orbit. Discussions of surface charging in polar orbits can be found in papers by Daly and Rodgers (1993) and Martin, 1991.

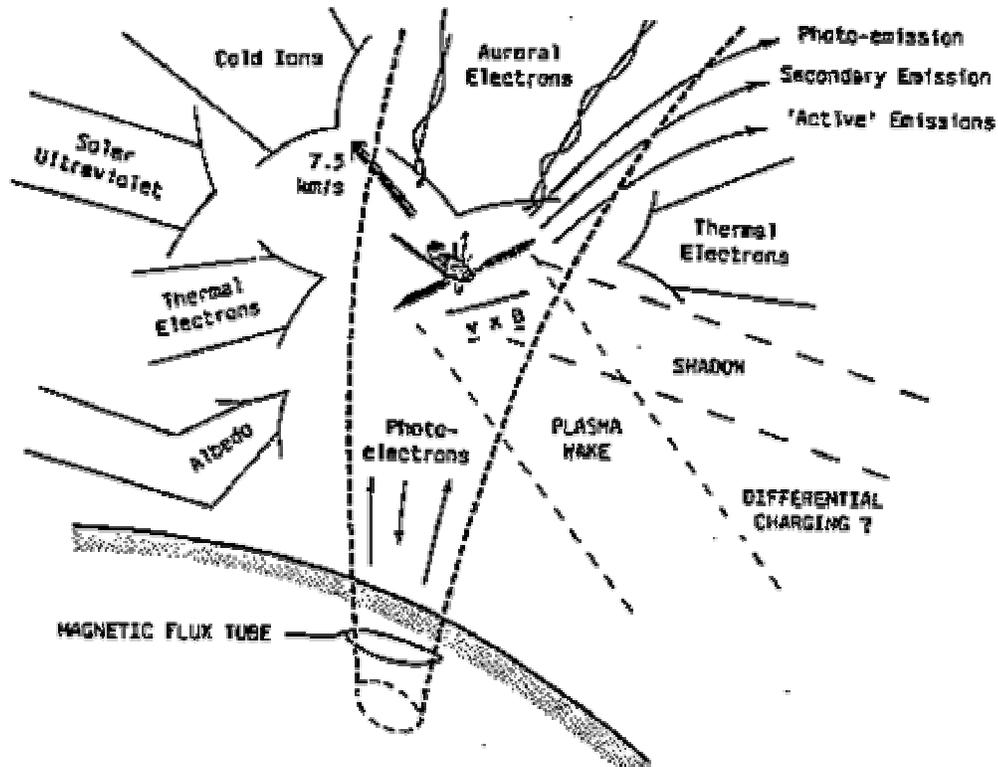

**Figure 7.** Plasma interactions with a spacecraft in polar orbit *(after Martin, 1991).*

### 3. Internal Charging

Internal charging, also called deep dielectric charging, is the buildup of electric charge inside the spacecraft due to the penetration of electrons with energies greater than 10 keV. The charge can accumulate on the surfaces or interiors of internal dielectric materials, or on the surfaces of insulated conductors inside the spacecraft. Internal charging depends on the orbital environment, the shielding thickness of the spacecraft, the geometry of the charged material, and the material properties, especially the conductivity of the material.

The rate at which the charge deposition inside the dielectric takes place determines the internal charging level (Garrett and Whittlesey, 1996). When this rate exceeds the rate at which charge can leak out, electric fields inside the dielectric increase in strength and eventually reach the breakdown strength of the dielectric material. When this happens, arc discharge and dielectric breakdown occur. A discharge is a fast transfer of charges which gives rise to transient currents. A typical voltage value across a dielectric leading to discharge is -10000 V. Discharges present serious hazards to spacecraft operations. They may damage electronic systems or lead to failure of various spacecraft components that are crucial for the proper functioning of the spacecraft.

## 4. Effects of Spacecraft Charging

The most important effect of spacecraft charging is the resulting electrostatic discharge (ESD). ESD can be in the form of surface discharge or in the form of bulk discharge (Estienne, 1993). A surface discharge occurs when the surface voltage exceeds the breakdown voltage of the surface material and as a result it can generate currents up to a few hundred amps. On the other hand, dielectric discharge is triggered when dielectrics are exposed to space radiation. The charge involved in bulk discharge is small relative to surface discharge, but nevertheless presents a direct hazard to electronics.

Arc discharges result mainly from differential charging and internal charging of spacecraft (Romero and Levy, 1993). Discharges may lead to anomalies such as erroneous logic changes in semiconductor devices, command errors or component failures. Degradation of sensors and solar cell panels is also a serious possibility and it may cause decreased amounts of power generation.

Discharges may also cause serious physical damage to surfaces. Localized heating and material loss result from arc discharges. Material loss may cause structural damage to the spacecraft. In addition, surface contamination can alter and degrade the properties of the surface materials.

The three types of discharges that can occur are "flashover", "punch-through" and "discharge to space". Flashover is the term given to the discharge from one surface to another. Punch-through is a discharge from the interior structure of a spacecraft through its surface, while discharge to space is the discharge from spacecraft to the surrounding plasma (Romero and Levy, 1993).

## 5. Computer Simulations and Design Guidelines

The purpose of the previous sections was to present to the reader a general overview of the basics involved in the analysis of spacecraft charging. Spacecraft charging is a major consideration in the design of space systems. In order to help engineers simulate charging effects on spacecraft, numerous computer simulation codes have been developed. An example of such a computer code is NASCAP- NASA Charging Analyzer Program (Katz et al, 1979; Roche and Purvis, 1979). NASCAP can simulate spacecraft charging for complex geometries and in various orbits (NASCAP-LEO, NASCAP-GEO). Furthermore, design guidelines have been documented to serve as references for spacecraft engineers. Although spacecraft design guidelines address the issue of surface charging extensively, internal charging is more of a challenge to simulate and hence the guidelines for controlling and preventing internal charging have not yet been extensively addressed. The reader is referred to the NASA Technical Paper 2361 (Purvis et al., 1984) and MIL-STD 1541A (Anon., 1987).

## IV. RADIATION HAZARDS TO ELECTRONICS IN SPACE

The harsh space radiation environment can have damaging effects on spacecraft electronics which may ultimately lead to mission failures. It is necessary to understand the space radiation environment in order to take measures for preventing radiation hazards by designing systems capable of withstanding radiation. This section of the paper is devoted to a discussion of the major sources of natural environmental radiation, as well as some basic damages they can produce in microelectronics.

The space radiation environment can be classified into two types: the trapped radiation environment and the non-trapped or transiting radiation environment (Stassinopoulos and Raymond, 1988). The trapped radiation environment includes the Van Allen radiation belts and consists of electrons, protons and some heavy ions and is influenced greatly by solar terrestrial interactions and the South Atlantic Anomaly (SAA). The transient radiation environment consists mainly of solar and galactic cosmic rays and solar flare particles, each of which contains high-energy protons and heavy ions. In the following subsections further elaboration on each of these radiation sources is presented. Moreover, some radiation damages such as ionization dose, atomic displacements and single-event phenomena are discussed.

### 1. The Trapped Radiation Environment

#### a. *Van Allen Radiation Belts*

The Van Allen radiation belts consist mainly of electrons and protons. They originate from the decay of neutrons produced by the interaction of cosmic rays with the low altitude atmospheric particles. These neutrons decay to yield protons, electrons and neutrinos. The neutrinos are almost massless and quickly escape into the cosmos, while the electrons and the protons are magnetically trapped by the Earth's magnetic field in regions called Van Allen belts. The trapped radiation of the Van Allen belts may also result from the acceleration of particles by magnetic storm activities or from solar flares (Alexander et al., 1994).

The motion of trapped charged particles is illustrated in Figure 8 (Spjeldvik and Rothwell, 1983). The particles sliding along the magnetic field lines slow down as they encounter strong magnetic fields, i.e. they are repelled from strong field regions and attracted towards low field regions. In addition to the spiraling motion, particles also bounce and drift around the Earth. Electrons drift eastward, while protons drift westward. The net effect is the formation of a ring current circulating westward.

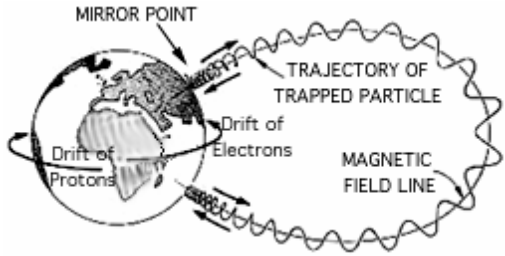

**Figure 8.** Trapped charged particle motions *(after Stern and Ness, 1981).*

Figure 9 (Strassinopoulos and Raymond, 1988) shows the electron and proton distributions in different regions of the magnetosphere. As shown in that figure, the energetic Van Allen belt electrons exist in two distinct regions: the "inner zone" and the "outer zone". The boundary between the two regions is at about 2.8 Earth radii. The inner zone electrons are less energetic than the outer zone electrons. While the inner zone electrons have energies < 5 MeV, the outer zone electron energies are around 7 MeV. The inner zone regime is relevant to LEO while the outer zone is relevant to the GEO environment.

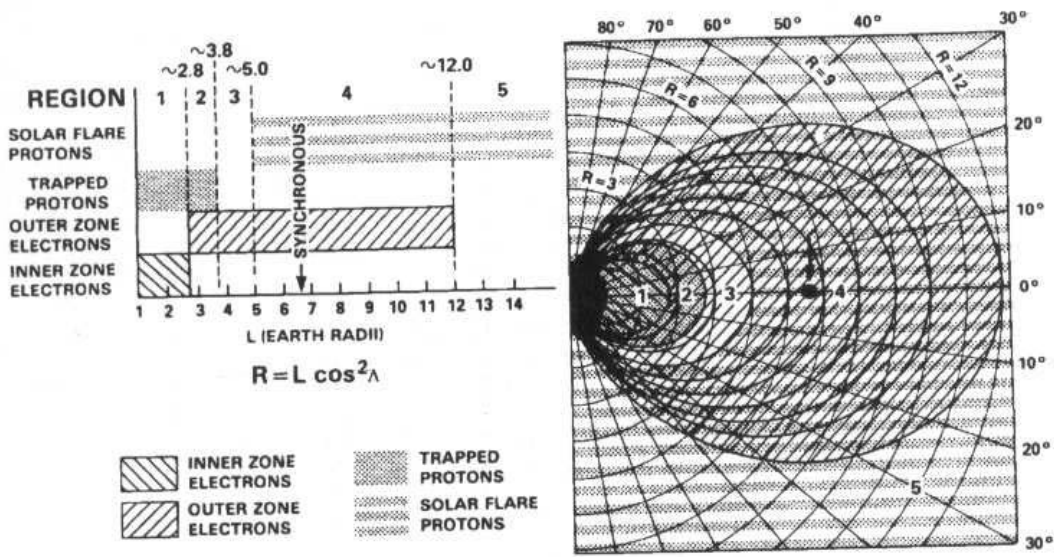

**Figure 9.** Distribution of charged particles in the magnetosphere *(after Stassinopoulos and Raymond, 1988).*

Figure 9 also shows proton distributions. In contrast to electrons, the protons cannot be divided into inner and outer zones since the most energetic protons are concentrated closer to the Earth as shown in Figure 10 (Strassinopoulos and Raymond, 1988). Trapped protons with energies > 10 MeV occupy regions 1 and 2 of Figure 9. Region 2 extends up to ~ 3.8 Earth radii.

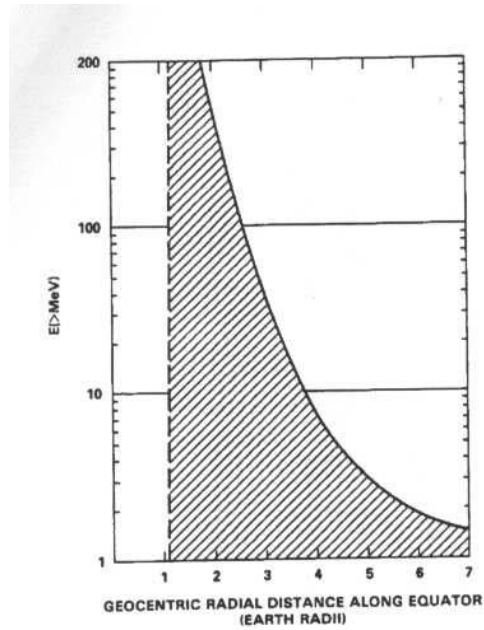

**Figure 10.** Trapped proton distribution as a function of energy *(after Stassinopoulos and Raymond, 1988).*

Figure 11 shows the inner and outer electron zones and the electron and proton densities (Anspaugh et al., 1982; Cladis, 1971).

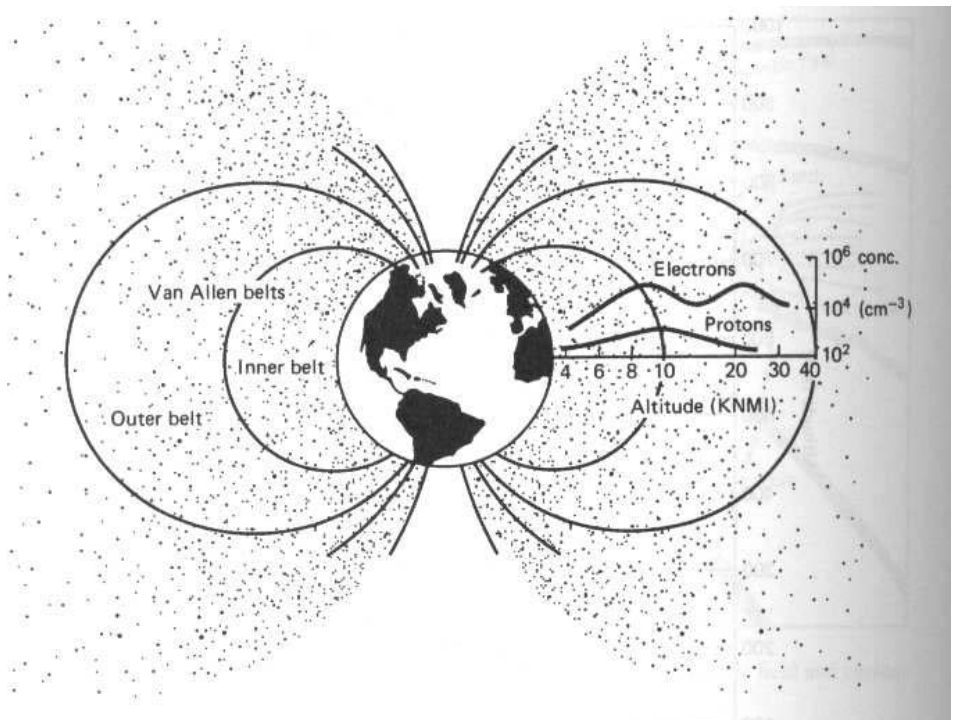

**Figure 11.** The Van Allen radiation belts showing electron and proton densities *(after Anspaugh et al., 1982; Cladis, 1971).*

The trapped particle environment is dynamic because variations in geomagnetic activity or solar cycle may influence particle fluxes. For instance, substorms result in the injection of energetic electrons into the midnight region, and these electrons drift towards dawn as mentioned earlier. Electron and proton intensities also vary with variations in the solar cycle. During solar minimum, the electron intensities decrease while proton intensities increase. In contrast, during solar maximum, electron intensities increase while proton intensities decrease (Stassinopoulos and Raymond, 1988).

*b. South Atlantic Anomaly*

As mentioned briefly in the first part of this paper, the Earth's magnetic field is a dipole which is offset from the Earth's center by ~ 500 km towards the Western Pacific, with the dipole axis inclined at about $11.5^o$ with respect to the Earth's rotation axis. This configuration results in an anomalous dip in the Earth's magnetic field over Brazil, where the magnetic field lines reenter the Earth. This property of the Earth's magnetic field is termed the South Atlantic Anomaly: SAA (see Figure 12).

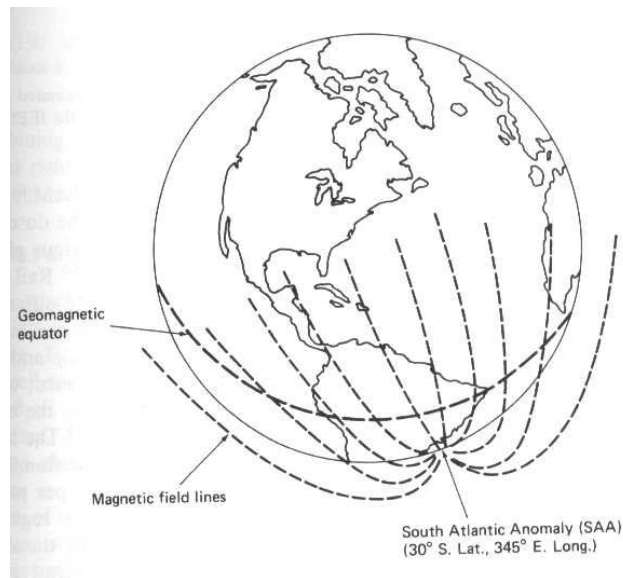

**Figure 12.** South Atlantic Anomaly *(after Messenger and Ash, 1991).*

Figure 13 below shows contour plots of the geomagnetic field at sea level. Since the SAA is a region of low magnetic field, the trapped radiation of the Van Allen belts reaches down to very low altitudes over the SAA. Therefore, being a region of high density trapped particles, the SAA presents a significant hazard to electronics of spacecraft in low altitude, low inclination orbits. In LEO, the most intense radiation is due to energetic protons in the South Atlantic Anomaly. Moreover, on the side of the Earth opposite to the SAA, there is an increase in the magnetic field and this region is called the Southeast-Asian Anomaly. The Southeast-Asian Anomaly expels the Van Allen radiation belts and pushes them to higher altitudes (Stassinopoulos and Raymond, 1988).

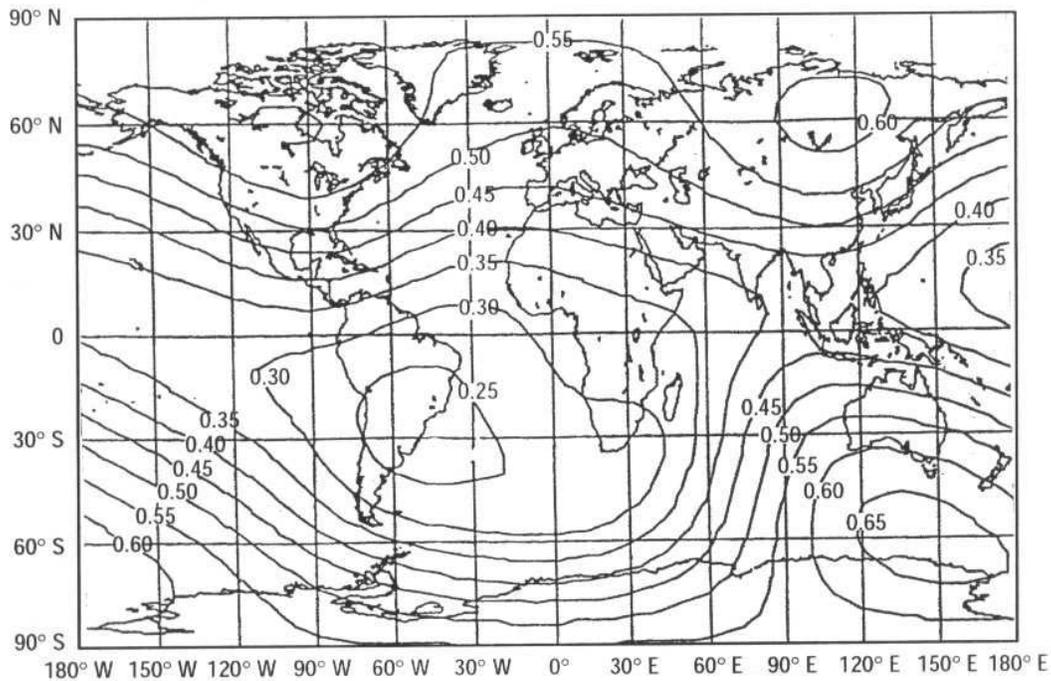

**Figure 13.** Contour plots of the geomagnetic field at sea level *(after NASA RP 1350).*

## 2. The Transient Radiation Environment

### a. Solar Cosmic Rays

Solar cosmic rays include solar energetic particles (SEP), which consist mainly of solar flare protons. A minority of alpha particles, heavy ions and electrons are also emitted. The flux of heavy ions is usually low. However, during increased solar events the ion flux may increase, and hence it can cause considerable damage to space electronics.

### b. Galactic Cosmic Rays

Galactic cosmic rays (GCR) originate from the far reaches of the galaxy. They contain ~ 85% protons, ~ 14% alpha particles and ~ 1% heavy nuclei. The galactic cosmic rays are energetic and can have energies in the order of GeV/nucleon. Therefore, they are capable of penetrating deep into semiconductor devices.

### c. Geomagnetic Field Effect

The geomagnetic field prevents the effect of solar and galactic cosmic rays on spacecraft orbiting in the low inclination LEO environment. This shielding is effective for inclinations up to 45°. Figure 14 (Stassinopoulos and Raymond, 1988) shows the ion and the Hydrogen energies required for particles to penetrate the magnetosphere at various altitudes.

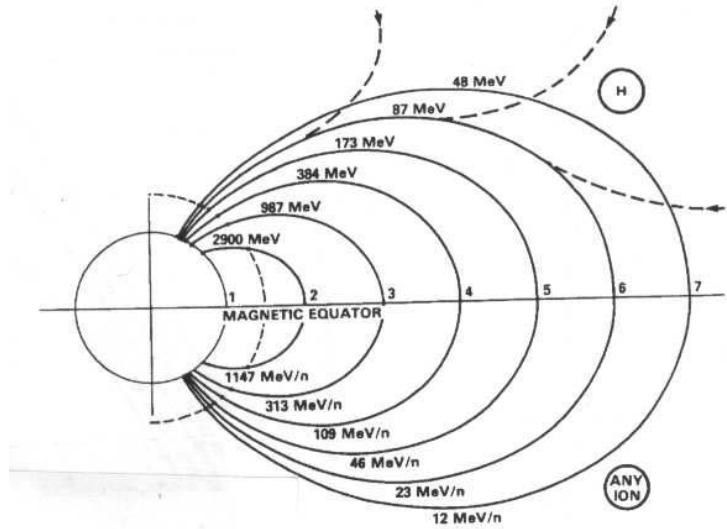

**Figure 14.** Hydrogen and ion energies required to penetrate the magnetosphere *(after Stassinopoulos and Raymond, 1988).*

On the other hand, geomagnetic shielding is not very effective for GEO spacecraft; thus energetic cosmic rays can penetrate the spacecraft surface.

## 3. Radiation Effects on Spacecraft Electronics

The radiation sources discussed are hazardous to electronics since energetic particles can deposit energy inside microelectronic circuitry and disrupt their proper operation. Energy deposition in electronics is measured in rads(M) where M is a specific material being considered (1 rad = 100 ergs/gm). Energy deposition can be in the form of ionization or atomic displacements, which can permanently damage electronics, or it can be in the form of single events, which can cause transient or permanent damages depending on the severity of the event. Electrons usually deposit energy by ionization, while energy deposition by protons can be either by ionization or by atomic displacements (Petersen, 1981). The following is only brief descriptions of the basic effects of radiation. Detailed descriptions of ionization effects, displacement effects and single event phenomena in various electronic circuits can be found in papers by Srour and McGarrity (1988), and Stephen (1993). The book by Messenger and Ash (1991) is a detailed treatment of radiation effects on electronic systems and includes guidelines for designing hardened electronic systems.

### a. *Total Ionizing Dose Effects*

The Total Ionizing Dose (TID) effect is the cumulative energy deposition in a material, which leads to degradation of electronics in the long run. TID limits the operational lifetime of spacecraft electronics. The main radiation sources of ionization effects are trapped electrons and protons of the Van Allen belts and solar flare protons. A common ionization effect is the gradual shift in the parameters of electronic components leading to circuit failures. For very sensitive microcircuits, ~1000 rads(Si) is enough to cause circuit failure. For hardened electronics, the failure dose can be as high as 10 Mega-rads(Si) (Raymond and Petersen, 1987). Shielding is usually used to reduce the ionization dose. Aluminum shields can effectively attenuate electrons and low-energy protons. However, high-energy protons (> 30 MeV) cannot be shielded.

### b. *Displacement Damage Effects*

Displacement damage refers to the displacement of atoms from their normal lattice positions. The absence of an atom from its lattice position gives rise to a *vacancy*, which is a radiation-induced defect. When the displaced atom moves to a non-lattice position, the resulting defect is termed an *interstitial*. Defect clusters are regions of disorder in the material, which result from multiple cumulative displacements. Defect clusters are sufficient to alter device properties and hence its performance (Srour and McGarrity, 1988). The main sources of displacement damage are energetic protons. Electrons usually do not produce significant displacements. Atomic displacements can seriously degrade solar cells. Displacement damages increase the resistance of solar cells, which in turn reduces the solar cell power generation.

### c. *Single Event Effects*

A Single event effect (SEE) is an anomaly caused by a single energetic particle striking a device. The single particle impact gives rise to an ionized track of electron-hole pairs along the particle's trajectory through a semiconductor material. Single event phenomena can be classified into various types (Kolasinski, 1989):

- Single Event Upset (SEU): A particle incident on a digital device causes an undesirable change in the logic state of the device. This is a transient effect that causes temporary problems. The device can be restored back to its original logic state (Liemohn, 1984).
- Single Event Latchup (SEL): SEL is similar to SEU, except that once the logic state of a device is changed by an incident particle, the device cannot be restored back to its original state. SEL can lead to device burnout unless the current to the device is limited by turning off the power supply.
- Single Event Burnout (SEB): In this case, large currents destroy the device. SEB commonly occurs in power MOSFETs and is the most dangerous form of single event since it leads to permanent failure.

## V. CONCLUSIONS

The purpose of this paper is to give the reader a glimpse of the natural space environment in relation to spacecraft charging and hazards to electronics and to make the reader aware of the significance and impacts of charging and radiation environments on spacecraft. The brief history introduced at the beginning of the paper outlines major observations and development stages since the early realization that spacecraft charging and radiation hazards must be major considerations in the design of space systems.

The first part of this paper has addressed the most important space environmental factors leading to spacecraft charging. A background on plasmas was provided, with special emphasis on the differences between ionospheric and substorm plasmas. Energetic particles, solar radiation and magnetic field effects were discussed as relevant to spacecraft charging mechanisms. Following the discussion of the charging environment, the basic mechanism of spacecraft surface charging which is defined by the current balance equation was introduced. Absolute, differential, and internal charging were also discussed. Differential and internal charging are the more dangerous forms of charging. Since the natural space environment vastly differs with altitude and latitude, the cases of surface charging in high altitude orbits, low altitude and latitude orbits and low altitude polar orbits were treated differently. The key point made there is that the level of surface charging in low altitude and latitude orbits is not significant when compared to the high level charging of surfaces in GEO and LEO polar orbits. Electrostatic discharge was introduced as the major damaging effect resulting from spacecraft charging. The importance of computer simulation tools and design guidelines were stressed.

The second part of this paper has addressed the space radiation environment, which is very rich in energetic electrons and protons trapped in the Van Allen belts, as well as heavier ions, galactic cosmic rays and solar flare particles. The solar cycle and geomagnetic substorms induce variations in the radiation environment. A particularly important property of the geomagnetic field is the region called South Atlantic Anomaly, which enhances trapped particle densities. Following a discussion of the fundamental radiation sources, a brief section was devoted to a discussion of radiation effects on electronics, including radiation dose effects, displacement damages and single event phenomena.

Nowadays, with the utilization of very sensitive electronic systems onboard spacecraft, surface charging and radiation effects have become important more than ever. In order to ensure the success of future space missions, our knowledge of spacecraft charging and radiation effects on electronics must continuously evolve in order to keep pace with the rapid advances of technology.